%% file: main.tex
\documentclass[sigconf, screen]{acmart}

\AtBeginDocument{%
  \providecommand\BibTeX{{%
    \normalfont B\kern-0.5em{\scshape i\kern-0.25em b}\kern-0.8em\TeX}}}

\setcopyright{acmlicensed}
\copyrightyear{2024}
\acmYear{2024}
\acmDOI{XXXXXXX.XXXXXXX}

\acmConference[Conference acronym 'XX]{Make sure to enter the correct
  conference title from your rights confirmation email}{June 03--05,
  2018}{Woodstock, NY}
\acmBooktitle{Woodstock '18: ACM Symposium on Neural Gaze Detection,
 June 03--05, 2018, Woodstock, NY} 
\acmISBN{978-1-4503-XXXX-X/18/06}

\input{preambles}

\copyrightyear{2024}
\acmYear{2024}
\setcopyright{rightsretained}
\acmConference[SA Technical Communications '24]{SIGGRAPH Asia 2024 Technical Communications}{2024 年 12 月 3 日～6 日}{東京、日本}
\acmBooktitle{SIGGRAPH Asia 2024 Technical Communications (SA Technical Communications '24), 2024 年 12 月 3 日～6 日、東京、日本}\acmDOI{10.1145/3681758. 3698020}
\acmISBN{979-8-4007-1140-4/24/ 12}

\begin{document}

\title{See-Through Face Display: Enabling Gaze Communication for Any Face---Human or AI}

\author{Kazuya Izumi}
\authornote{Both authors contributed equally to the paper}
\email{izumin@digitalnature.slis.tsukuba.ac.jp}
\affiliation{%
  \institution{University of Tsukuba}
  \country{Japan}
}

\author{Ryosuke Hyakuta}
\authornotemark[1]
\email{momosuke@digitalnature.slis.tsukuba.ac.jp}
\affiliation{%
  \institution{University of Tsukuba}
  \country{Japan}
}

\author{Ippei Suzuki}
\email{1heisuzuki@digitalnature.slis.tsukuba.ac.jp}
\affiliation{%
  \institution{University of Tsukuba}
  \country{Japan}
}

\author{Yoichi Ochiai}
\email{wizard@slis.tsukuba.ac.jp}
\affiliation{%
  \institution{\mbox{R\&D Center for Digital Nature}}
  \country{Japan}
}


\input{sections/00_abstract}

\ccsdesc[500]{Human-centered computing~Displays and imagers}

\keywords{Telepresence, Eye Contact, Gaze Awareness, Nonverbal Communication}

\begin{teaserfigure}
  \includegraphics[width=\textwidth]{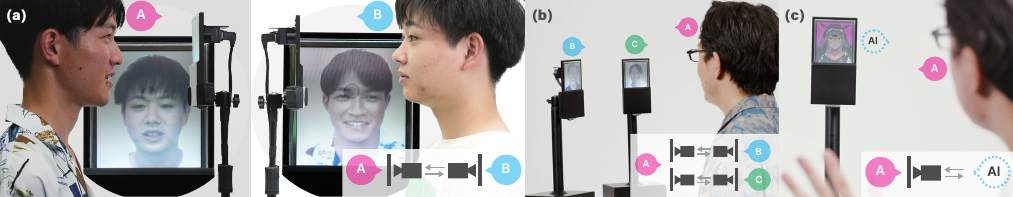}
    \caption{See-Through Face Display supports diverse face and gaze communication scenarios, extending beyond the scope of human-to-human or human-to-AI interactions. (a) One-on-one remote conversation. (b) Three-person remote conversation. (c) Nonverbal communication with an AI avatar.}
  \Description{teaser}
  \label{fig:teaser}
\end{teaserfigure}

\maketitle

\input{sections/01_introduction}
\input{sections/02_related_works}
\input{sections/03_implementation}
\input{sections/04_applications}
\input{sections/05_discussion}
\input{sections/06_conclusion}

\begin{acks}
We are grateful to Japan Display Inc. for lending us the prototype of See-Through Face Display and Kazuhiko Sako, Kazunari Tomizawa, and Kentaro Okuyama for their technical assistance in the hardware development.
\end{acks}

\bibliographystyle{ACM-Reference-Format}
\bibliography{reference}

\end{document}

%% file: preambles.tex
\usepackage[whole]{bxcjkjatype}
\usepackage[unicode]{hyperref}

\usepackage{xcolor}
\usepackage{graphicx}

\usepackage{amsmath}
\usepackage{caption}
\usepackage{comment}
\usepackage{docmute}
\usepackage{empheq}
\usepackage{enumitem}
\usepackage{here}
\usepackage{karnaugh-map}
\usepackage{lscape}
\usepackage{multirow}
\usepackage{setspace}
\usepackage{siunitx}
\usepackage{url}

\usepackage{twemojis}
\usepackage{array}
\usepackage{amssymb} 
\usepackage{pifont}  
\usepackage{xcolor}  
\usepackage{algorithm}
\usepackage{algorithmic}
\usepackage{booktabs}
\usepackage{subcaption}
\usepackage[english]{babel}  
\usepackage{graphicx}  

\definecolor{ForestGreen}{RGB}{34,139,34}


%% file: sections/00_abstract.tex
\begin{abstract}

We present See-Through Face Display, an eye-contact display system designed to enhance gaze awareness in both human-to-human and human-to-avatar communication. The system addresses the limitations of existing gaze correction methods by combining a transparent display with a strategically positioned camera. The display alternates rapidly between a visible and transparent state, thereby enabling the camera to capture clear images of the user's face from behind the display. This configuration allows for mutual gaze awareness among remote participants without the necessity of a large form factor or computationally resource-intensive image processing.
In comparison to conventional methodologies, See-Through Face Display offers a number of practical advantages. The system requires minimal physical space, operates with low computational overhead, and avoids the visual artifacts typically associated with software-based gaze redirection. These features render the system suitable for a variety of applications, including multi-party teleconferencing and remote customer service. Furthermore, the alignment of the camera's field of view with the displayed face position facilitates more natural gaze-based interactions with AI avatars. This paper presents the implementation of See-Through Face Display and examines its potential applications, demonstrating how this compact eye-contact system can enhance gaze communication in both human-to-human and human-AI interactions.

\end{abstract}

%% file: sections/01_introduction.tex
\section{Introduction}

Building on the ClearBoard concept~\cite{Ishii1992-gq}, gaze awareness in telepresence has become a key topic in Human-Computer Interaction.
Subsequently, researchers have proposed various methods to address gaze awareness in different contexts, including multi-party teleconferencing systems~\cite{Okada1994-oa, Otsuka2016-wx} and remote instructions~\cite{Faridan2023-bc}.

However, conventional methods of gaze correction have several limitations that restrict their practical everyday use.
Many optical gaze redirection methods require large form factors, which significantly constrains the installation locations.
While image processing-based gaze redirection systems require small form factors, they often encounter challenges such as high computational resource requirements and visual artifacts that diminish the user experience.
As the application area of telepresence systems expands, the demand for compact, resource-efficient eye contact systems that provide a high-quality user experience is increasing.

In order to address these challenges, we propose ``See-Through Face Display,'' an eye-contact display system composed of a transparent display and a camera situated behind the display.
The transparent display alternates rapidly between a visible and a transparent state.
When the display is transparent state, the camera captures images of the user’s face from behind the display.
Consequently, when the local user gazes at the remote user's face displayed on the screen, the local user's gaze is aligned with the optical axis of the camera, thereby enabling the remote user to perceive eye contact with the local user.

See-Through Face Display offers a number of practical improvements over existing gaze correction methods.
In contrast to conventional optical methods, this display reduces the physical hardware requirements.
In contrast to image processing-based methods, this display performs gaze correction with minimal computational overhead and avoids visual artifacts.
These features make See-Through Face Display an effective solution for a wide range of gaze communication applications.
Moreover, the alignment of the camera's field of view with the displayed face position allows for the extension of human-to-human gaze communication to encompass non-verbal interactions with AI avatars.

In this paper, we outline the implementation of See-Through Face Display and discuss its future applications, exploring how this compact eye contact display can serve as a portal for everyday gaze communication with both humans and AI.

%% file: sections/02_related_works.tex
\section{Related Work}

\subsection{Gaze Awareness in Telepresence}

In the field of telepresence, two main approaches have been developed to improve user gaze awareness: optical approaches that align the displayed user's facial position with the camera's optical axis~\cite{Ishii1992-gq, Okada1994-oa, Otsuka2016-wx}, and software approaches that redirect the user's gaze using image processing~\cite{Wood2018-bq, wang2021facevid2vid}.
The optical approach emerged with ClearBoard~\cite{Ishii1992-gq}, which focused on one-to-one remote collaboration.
Subsequently, some research such as MAJIC~\cite{Okada1994-oa} and MMSpace~\cite{Otsuka2016-wx} extended the concept to multi-party conferencing.
However, these systems tend to have very large form factors, limiting use outside of specific applications. 
On the other hand, software approaches typically detect the user's eye region in captured images and manipulate the pupil position to redirect gaze.
While this method can potentially solve the form factor issue, it often requires significant computational resources for real-time image processing.
In addition, these approaches have the potential of causing visual artifacts that can affect the user experience.
See-Through Face Display offers advantages over both approaches.
It implements a more compact form factor compared to previous optical approaches, while avoiding the high computational requirements and potential visual artifacts associated with software approaches.
This makes our system suitable for a wider range of applications, potentially supporting everyday communication between people at a distance.

\subsection{Interactions Using See-Through Displays}

See-through displays have been proposed and used in various human-computer interaction scenarios. These displays offer unique opportunities for creating interaction paradigms and enhancing the user experience.
SecondLight~\cite{Lindlbauer2014-il} presented an approach using a projector to display images on a screen that could rapidly switch between diffuse and clear states. The system incorporated a camera positioned behind the screen to capture the environment, enabling interactions such as multi-touch and tangible surface interactions.
Tracs~\cite{Izadi2008-xg} implemented an LCD screen with locally controllable transparency. This technology enabled face-to-face collaboration between users on opposite sides of the screen, while maintaining eye contact and privacy of screen content.

See-through Face Display is particularly focused on applications for telepresence systems.
This is made possible by the display's capacity to provide sufficient performance for telepresence systems, in terms of display brightness and resolution, and input video frame rate and resolution.
See-Through Face Display is also more compact than previous research.
Instead of using projectors or complex optical arrangements, it uses an LED LCD panel 
This design choice has the effect of reducing the system's form factor and computational resource requirement, thereby rendering it more suitable for everyday use in a variety of environments.

%% file: sections/03_implementation.tex
\section{System}

\subsection{Hardware Setup}

See-Through Face Display consists of a transparent display and a camera situated behind the display as shown in \autoref{fig:hardware}(a).
We use a 4-inch full-color transparent LCD manufactured by Japan Display Inc. with a resolution of $320\times360$ pixels~\cite{8006061, okuyama202138}.
This display attains transparency by rapidly alternating between transparent and scattering states within the liquid crystal layer, as shown in \autoref{fig:hardware}(b). 
The camera situated behind the display receives signals from the driver circuit of the transparent display to capture images.
The driver circuit transmits capture signals in the transparent state, enabling the display to show images from the HDMI signal and to capture the user's face behind the transparent display.
We use Toshiba Teli Corporation's BU160MCF for the camera, which captures images based on signals from the transparent display.
The capture time is set at 6 ms within a frame (60 FPS; 16.7 ms).
The above system enables the user's gaze towards the conversation partner shown on the display to align closely with the optical axis of the camera, facilitating eye contact with the user depicted in the captured video image.

\subsection{Image Processing}

\begin{figure}[t]
  \centering
  \includegraphics[width=\linewidth]{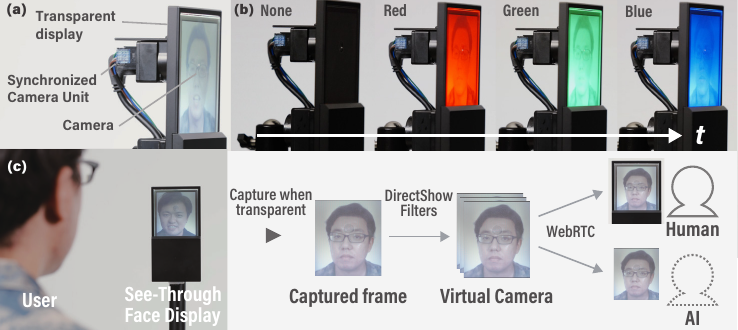}
    \caption{(a) Hardware setup of See-Through Face Display. (b) The transparent display rapidly switches between transparent (None) and scattering (Red, Green, Blue). When the display is transparent, the camera captures the user's face. (c) Image Processing of See-Through Face Display.
    The captured frames are streamed to a virtual camera using DirectShow Filters and transmitted via the WebRTC protocol.}
    \label{fig:hardware}
\end{figure}

The captured image of the user's face is processed and transmitted to a human or AI, as shown in \autoref{fig:hardware}(c).
The image captured by the camera module is initially processed through a DirectShow Filters\footnote{\url{https://learn.microsoft.com/en-us/windows/win32/directshow/directshow-filters}}.
This processed video stream is then made available to applications as a virtual camera device.
The video and audio of the user captured by the camera is sent to the remote system via a peer-to-peer connection using the WebRTC protocol.
In conversations involving two or more users, each participant can set up multiple displays with cameras to accommodate the number of participants and establish individual connections.

\begin{figure*}[t]
  \centering
  \includegraphics[width=\linewidth]{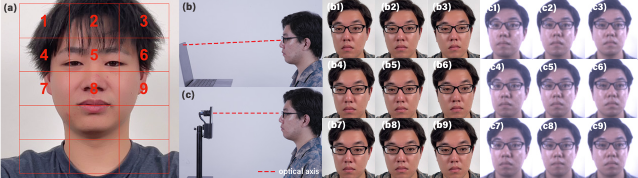}
  \caption{Evaluation of Eye Contact Comparison. (a) A participant watched a picture of a person's face and a nine-part grid of areas near the face as stimulation. (b) Eye contact evaluation when watching the stimulation with a conventional videoconferencing system. (c) Eye contact evaluation when watching the stimulation with See-Through Face Display.}
\label{fig:evaluation}
\end{figure*}

In our evaluation, a participant was shown with \autoref{fig:evaluation}(a), via either a conventional laptop-based video conferencing configuration (\autoref{fig:evaluation}(b)) or See-Through Face Display (\autoref{fig:evaluation}(c)).
The participant were instructed to direct their gaze at nine specific regions of the reference image (labeled 1-9).
As shown in autoref{fig:evaluation}(b1-b9), the participant in the conventional videoconferencing configuration appeared to be directing the gaze slightly downward, suggesting that eye contact could not have occurred.
In contrast, as shown in \autoref{fig:evaluation}(c1-c9), See-Through Face Display improves gaze awareness and demonstrates that eye contact is correctly occurring when the participant gazes at the areas close to the eyes, particularly in regions 2 and 5.

%% file: sections/04_applications.tex
\section{Application Scenarios}

In this section, we present several potential application scenarios of See-Through Face Display that demonstrate the versatility and practicality of this technology.

\subsection{Multi-Party Remote Conversations}
One of the most typical applications of See-Through Face Display is the facilitation of multi-party remote conversations. In this scenario, the ability of participants to clearly see one another enhances the natural flow of conversation.
While previous multi-party telepresence systems have offered similar capabilities, this display's compact form factor allows for its use in environments with limited space, such as restaurants, where larger setups would be impractical.
This has the potential to expand the range of locations in which rich, gaze-aware remote interactions can take place.
In addition, this display is expected more effective in situations where eye contact is more important, such as Deaf people's sign language conversations.

\subsection{Remote Reception Services}
In recent years, there has been a notable increase in the utilization of remote reception services, wherein store personnel engage with customers from a distance.
Non-verbal communication plays a critical role in effective customer service and remote guidance.
The display enables seamless communication of nonverbal cues between remote staff and local customers, such as beginning to talk through eye contact.
Furthermore, See-Through Face Display facilitates the incorporation of camera-based procedures in a seamless manner.
For example, if a customer requires the scanning of a Quick Response (QR) code displayed on their mobile device, they are able to do so via the display without disrupting the face-to-face interaction with the remote staff member.

\subsection{AI-Assisted Product Recommendations}
\begin{figure}[t]
  \centering
  \includegraphics[width=\linewidth]{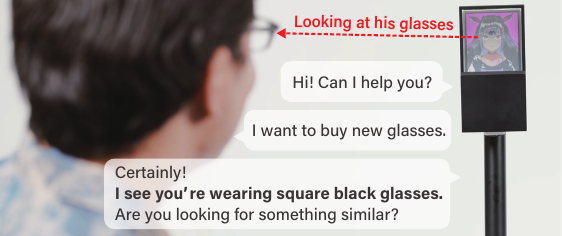}
        \caption{An application scenario for AI-assisted product recommendations. When a visitor gazes at the avatar, the avatar talks to him and offers recommendations based on his requests and his current glasses while looking at him.}
    \label{fig:applications}
\end{figure}

See-Through Face Display enables the camera's field of view to be aligned with the displayed face position, thereby facilitating interactions not only with humans but also with AI avatars.
For example, one potential application is the installation of the display in a store, which would enable an AI avatar to make product recommendations, as shown in \autoref{fig:applications}.
When a visitor directs their gaze toward the AI avatar, the avatar collects information about the visitor and the surrounding environment through the camera while actually looking at the visitor.
Based on this information and the customer's requests, the AI can present personalized information and recommendations.
This type of casual, everyday gaze communication with AI becomes possible through this display, leveraging its low computational requirements and high real-time performance.
It introduces a new approach for human-AI interaction in public spaces, making such interactions more natural and intuitive.

%% file: sections/05_discussion.tex
\section{Discussion}

See-Through Face Display enables gaze communication between humans and AI through a display with a small form factor, thereby facilitating the exploration of interaction paradigms.
Although the potential for gaze awareness has been demonstrated, as shown in \autoref{fig:evaluation}, a comprehensive technical evaluation has yet to be conducted.
Research indicates that humans typically perceive lack of eye contact when gaze deviates by more than five degrees from the center~\cite{Jaklic2017-dm}.
Building on this understanding, future studies will employ more precise measurements.

In addition, as shown in \autoref{fig:evaluation}(b), the image quality and contrast of the video captured from the see-through face display are also problematic. This is due to the fact that the user's face is captured only when the display module is transparent, and we will improve the performance in the future.
Although evaluation of eye contact correction using this display has already been conducted, further comprehensive user studies are required.
These future investigations will focus on evaluating users' impressions of both human participants and AI avatars when interacting through See-Through Face Display.
The purpose of this study is to provide insights into the effectiveness of this display in creating engaging and natural-feeling gaze interactions.

%% file: sections/06_conclusion.tex
\section{Conclusion}

See-Through Face Display is a compact and efficient system designed to enhance gaze awareness with any face.
The integration of a transparent display with a strategically positioned camera addresses the limitations of previous telepresence systems, enabling mutual gaze communication with minimal computational resources and without visual artifacts.
The system's versatility has been demonstrated across various application scenarios, including multi-party conversations and AI interactions. Its capability to facilitate natural eye contact in diverse settings suggests the potential for integrating gaze-aware communication into daily life.
As remote interactions become increasingly common, this display offers a promising advancement toward more natural and engaging virtual communications, which are not limited to human or AI interactions.

%% file: main.bbl

\begin{thebibliography}{11}


\ifx \showCODEN    \undefined \def \showCODEN     #1{\unskip}     \fi
\ifx \showDOI      \undefined \def \showDOI       #1{#1}\fi
\ifx \showISBNx    \undefined \def \showISBNx     #1{\unskip}     \fi
\ifx \showISBNxiii \undefined \def \showISBNxiii  #1{\unskip}     \fi
\ifx \showISSN     \undefined \def \showISSN      #1{\unskip}     \fi
\ifx \showLCCN     \undefined \def \showLCCN      #1{\unskip}     \fi
\ifx \shownote     \undefined \def \shownote      #1{#1}          \fi
\ifx \showarticletitle \undefined \def \showarticletitle #1{#1}   \fi
\ifx \showURL      \undefined \def \showURL       {\relax}        \fi
\providecommand\bibfield[2]{#2}
\providecommand\bibinfo[2]{#2}
\providecommand\natexlab[1]{#1}
\providecommand\showeprint[2][]{arXiv:#2}

\bibitem[Faridan et~al\mbox{.}(2023)]%
        {Faridan2023-bc}
\bibfield{author}{\bibinfo{person}{Mehrad Faridan}, \bibinfo{person}{Bheesha Kumari}, {and} \bibinfo{person}{Ryo Suzuki}.} \bibinfo{year}{2023}\natexlab{}.
\newblock \showarticletitle{{ChameleonControl: Teleoperating real human surrogates through mixed reality gestural guidance for remote hands-on classrooms}}. In \bibinfo{booktitle}{\emph{{Proceedings of the 2023 CHI Conference on Human Factors in Computing Systems}}}. \bibinfo{publisher}{ACM}, \bibinfo{address}{New York, NY, USA}, \bibinfo{pages}{1--13}.
\newblock
\urldef\tempurl%
\url{https://doi.org/10.1145/3544548.3581381}
\showDOI{\tempurl}


\bibitem[Ishii and Kobayashi(1992)]%
        {Ishii1992-gq}
\bibfield{author}{\bibinfo{person}{Hiroshi Ishii} {and} \bibinfo{person}{Minoru Kobayashi}.} \bibinfo{year}{1992}\natexlab{}.
\newblock \showarticletitle{{ClearBoard}: a seamless medium for shared drawing and conversation with eye contact}. In \bibinfo{booktitle}{\emph{Proceedings of the {SIGCHI} Conference on Human Factors in Computing Systems}} (Monterey, California, USA) \emph{(\bibinfo{series}{CHI '92})}. \bibinfo{publisher}{Association for Computing Machinery}, \bibinfo{address}{New York, NY, USA}, \bibinfo{pages}{525--532}.
\newblock
\showISBNx{9780897915137}
\urldef\tempurl%
\url{https://doi.org/10.1145/142750.142977}
\showDOI{\tempurl}


\bibitem[Izadi et~al\mbox{.}(2008)]%
        {Izadi2008-xg}
\bibfield{author}{\bibinfo{person}{Shahram Izadi}, \bibinfo{person}{Steve Hodges}, \bibinfo{person}{Stuart Taylor}, \bibinfo{person}{Dan Rosenfeld}, \bibinfo{person}{Nicolas Villar}, \bibinfo{person}{Alex Butler}, {and} \bibinfo{person}{Jonathan Westhues}.} \bibinfo{year}{2008}\natexlab{}.
\newblock \showarticletitle{{Going beyond the display: a surface technology with an electronically switchable diffuser}}. In \bibinfo{booktitle}{\emph{{Proceedings of the 21st annual ACM symposium on User interface software and technology}}}. \bibinfo{publisher}{ACM}, \bibinfo{address}{New York, NY, USA}.
\newblock
\showISBNx{9781595939753}
\urldef\tempurl%
\url{https://doi.org/10.1145/1449715.1449760}
\showDOI{\tempurl}


\bibitem[Jakli{\v c} et~al\mbox{.}(2017)]%
        {Jaklic2017-dm}
\bibfield{author}{\bibinfo{person}{Ale{\v s} Jakli{\v c}}, \bibinfo{person}{Franc Solina}, {and} \bibinfo{person}{Luka {\v S}ajn}.} \bibinfo{year}{2017}\natexlab{}.
\newblock \showarticletitle{User interface for a better eye contact in videoconferencing}.
\newblock \bibinfo{journal}{\emph{Displays}}  \bibinfo{volume}{46} (\bibinfo{date}{Jan.} \bibinfo{year}{2017}), \bibinfo{pages}{25--36}.
\newblock
\showISSN{0141-9382}
\urldef\tempurl%
\url{https://doi.org/10.1016/j.displa.2016.12.002}
\showDOI{\tempurl}


\bibitem[Lindlbauer et~al\mbox{.}(2014)]%
        {Lindlbauer2014-il}
\bibfield{author}{\bibinfo{person}{David Lindlbauer}, \bibinfo{person}{Toru Aoki}, \bibinfo{person}{Robert Walter}, \bibinfo{person}{Yuji Uema}, \bibinfo{person}{Anita Höchtl}, \bibinfo{person}{Michael Haller}, \bibinfo{person}{Masahiko Inami}, {and} \bibinfo{person}{Jörg Müller}.} \bibinfo{year}{2014}\natexlab{}.
\newblock \showarticletitle{{Tracs: transparency-control for see-through displays}}. In \bibinfo{booktitle}{\emph{{Proceedings of the 27th annual ACM symposium on User interface software and technology}}}. \bibinfo{publisher}{ACM}, \bibinfo{address}{New York, NY, USA}.
\newblock
\showISBNx{9781450330695}
\urldef\tempurl%
\url{https://doi.org/10.1145/2642918.2647350}
\showDOI{\tempurl}


\bibitem[Okada et~al\mbox{.}(1994)]%
        {Okada1994-oa}
\bibfield{author}{\bibinfo{person}{Ken-Ichi Okada}, \bibinfo{person}{Fumihiko Maeda}, \bibinfo{person}{Yusuke Ichikawaa}, {and} \bibinfo{person}{Yutaka Matsushita}.} \bibinfo{year}{1994}\natexlab{}.
\newblock \showarticletitle{{Multiparty videoconferencing at virtual social distance: MAJIC design}}. In \bibinfo{booktitle}{\emph{{Proceedings of the 1994 ACM conference on Computer supported cooperative work}}} \emph{(\bibinfo{series}{CSCW '94})}. \bibinfo{publisher}{Association for Computing Machinery}, \bibinfo{address}{New York, NY, USA}, \bibinfo{pages}{385–393}.
\newblock
\showISBNx{9780897916899}
\urldef\tempurl%
\url{https://doi.org/10.1145/192844.193054}
\showDOI{\tempurl}


\bibitem[Okuyama et~al\mbox{.}(2017)]%
        {8006061}
\bibfield{author}{\bibinfo{person}{Kentaro Okuyama}, \bibinfo{person}{Tae Nakahara}, \bibinfo{person}{Yudai Numata}, \bibinfo{person}{Tenfu Nakamura}, \bibinfo{person}{Manabu Mizuno}, \bibinfo{person}{Hiroki Sugiyama}, \bibinfo{person}{Shinichiro Nomura}, \bibinfo{person}{Shunpei Takeuchi}, \bibinfo{person}{Yoshihide Oue}, \bibinfo{person}{Hirofumi Kato}, \bibinfo{person}{Shohei Ito}, \bibinfo{person}{Akira Hasegawa}, \bibinfo{person}{Tadafumi Ozaki}, \bibinfo{person}{Mamoru Douyou}, \bibinfo{person}{Takayuki Imai}, \bibinfo{person}{Keiji Takizawa}, {and} \bibinfo{person}{Satoshi Matsushima}.} \bibinfo{year}{2017}\natexlab{}.
\newblock \showarticletitle{{79‐4L: \textit{late‐news paper}: Highly transparent LCD using new scattering‐type liquid crystal with field sequential color edge light}}.
\newblock \bibinfo{journal}{\emph{Digest of technical papers. SID International Symposium}} \bibinfo{volume}{48}, \bibinfo{number}{1}, \bibinfo{pages}{1166--1169}.
\newblock
\showISSN{0097-966X,2168-0159}
\urldef\tempurl%
\url{https://doi.org/10.1002/sdtp.11851}
\showDOI{\tempurl}


\bibitem[Okuyama et~al\mbox{.}(2021)]%
        {okuyama202138}
\bibfield{author}{\bibinfo{person}{Kentaro Okuyama}, \bibinfo{person}{Yuji Omori}, \bibinfo{person}{Makoto Miyao}, \bibinfo{person}{Koji Kitamura}, \bibinfo{person}{Muneaki Zako}, \bibinfo{person}{Yoshio Maruoka}, \bibinfo{person}{Kenichi Akutsu}, \bibinfo{person}{Hiroki Sugiyama}, \bibinfo{person}{Yoshihide Oue}, \bibinfo{person}{Tenfu Nakamura}, {et~al\mbox{.}}} \bibinfo{year}{2021}\natexlab{}.
\newblock \showarticletitle{38-2: Invited Paper: 12.3-in Highly Transparent LCD by Scattering Mode with Direct Edge Light and Field-Sequential Color-Driving Method}. In \bibinfo{booktitle}{\emph{SID Symposium Digest of Technical Papers}}, Vol.~\bibinfo{volume}{52}. Wiley Online Library, \bibinfo{pages}{519--522}.
\newblock
\urldef\tempurl%
\url{https://doi.org/10.1002/sdtp.14732}
\showURL{%
\tempurl}


\bibitem[Otsuka(2016)]%
        {Otsuka2016-wx}
\bibfield{author}{\bibinfo{person}{Kazuhiro Otsuka}.} \bibinfo{year}{2016}\natexlab{}.
\newblock \showarticletitle{{MMSpace: Kinetically-augmented telepresence for small group-to-group conversations}}. In \bibinfo{booktitle}{\emph{{2016 IEEE virtual reality (VR)}}}. \bibinfo{publisher}{IEEE}, \bibinfo{address}{Greenville, SC, USA}, \bibinfo{pages}{19–28}.
\newblock
\showISBNx{9781509008360}
\urldef\tempurl%
\url{https://doi.org/10.1109/VR.2016.7504684}
\showDOI{\tempurl}


\bibitem[Wang et~al\mbox{.}(2021)]%
        {wang2021facevid2vid}
\bibfield{author}{\bibinfo{person}{Ting-Chun Wang}, \bibinfo{person}{Arun Mallya}, {and} \bibinfo{person}{Ming-Yu Liu}.} \bibinfo{year}{2021}\natexlab{}.
\newblock \showarticletitle{One-Shot Free-View Neural Talking-Head Synthesis for Video Conferencing}. In \bibinfo{booktitle}{\emph{Proceedings of the IEEE Conference on Computer Vision and Pattern Recognition}}.
\newblock


\bibitem[Wood et~al\mbox{.}(2018)]%
        {Wood2018-bq}
\bibfield{author}{\bibinfo{person}{Erroll Wood}, \bibinfo{person}{Tadas Baltru{\v s}aitis}, \bibinfo{person}{Louis-Philippe Morency}, \bibinfo{person}{Peter Robinson}, {and} \bibinfo{person}{Andreas Bulling}.} \bibinfo{year}{2018}\natexlab{}.
\newblock \showarticletitle{{GazeDirector}: Fully articulated eye gaze redirection in video}.
\newblock \bibinfo{journal}{\emph{Comput. Graph. Forum}} \bibinfo{volume}{37}, \bibinfo{number}{2} (\bibinfo{date}{May} \bibinfo{year}{2018}), \bibinfo{pages}{217--225}.
\newblock
\showISSN{0167-7055, 1467-8659}
\urldef\tempurl%
\url{https://doi.org/10.1111/cgf.13355}
\showDOI{\tempurl}


\end{thebibliography}
